\documentclass[prd,preprintnumbers,nofootinbib,floatfix,twocolumn,notitlepage]{revtex4-1}

\usepackage{textcase}
\usepackage{graphicx,epsfig,psfrag,amssymb,hyperref}
\usepackage{multirow}
\usepackage{color,graphicx,epsfig,psfrag,amsmath,empheq}
\usepackage{bm}
\usepackage{mathrsfs,amsfonts,, color}
\usepackage[caption=false]{subfig}
\usepackage{hepunits}

\newcommand{\tev}{\ensuremath{\mathrm{\: Te\kern -0.1em V}}\xspace}
\newcommand{\gev}{\ensuremath{\mathrm{\: Ge\kern -0.1em V}}\xspace}
\newcommand{\mev}{\ensuremath{\mathrm{\: Me\kern -0.1em V}}\xspace}

\begin{document}

\title{Photoproduction of leptophobic bosons}

\author{Cristiano Fanelli}
\email{cfanelli@mit.edu}

\author{Mike Williams}
\email{mwill@mit.edu}
\affiliation{Laboratory for Nuclear Science, Massachusetts Institute of Technology, Cambridge, MA 02139, U.S.A.}

\begin{abstract}
We propose a search for photoproduction of leptophobic bosons that couple to quarks at the {\sc GlueX} experiment at Jefferson Lab. 
We study in detail a new gauge boson that couples to baryon number $B$, and estimate that $\gamma p \to p B$ will provide the best sensitivity for $B$ masses above 0.5\gev. 
This search will also provide sensitivity to other proposed dark-sector states that couple to quarks. 
Finally, our results motivate a similar search for $B$ boson electroproduction at the CLAS experiment. 
\end{abstract}

\maketitle

\section{Introduction}

The possibility that dark matter may interact via as-yet-unknown forces 
has motivated a dedicated worldwide effort to search for dark forces  (see Ref.~\cite{Essig:2013lka} for a review).
Most existing and planned searches for dark bosons rely on leptonic couplings; therefore, the constraints on dark forces that couple predominantly to quarks are significantly weaker than on those with substantial couplings to leptons~\cite{Tulin:2014tya}. 
A compelling -- yet simple -- dark-force scenario introduces a new $U(1)_B$ gauge symmetry that couples to baryon number~\cite{Nelson:1989fx,Rajpoot:1989jb,Foot:1989ts,He:1989mi,Carone:1994aa,Bailey:1994qv,Carone:1995pu,Aranda:1998fr,FileviezPerez:2010gw,Graesser:2011vj}. Assuming the symmetry is broken gives rise to a massive vector boson referred to as the $B$ boson.  
It is well known that $U(1)_B$ is anomalous, and that additional fermions with electroweak quantum numbers are required to cancel the anomalies~\cite{Foot:1989ts}. 
Such massive fermions could be stable providing viable dark-matter candidates. 

Beyond this specific model, searches for $B$ bosons are sensitive to so-called dark $\rho$ mesons and other proposed dark-sector states that mix with QCD vector mesons (see, {\em e.g.}, Ref.~\cite{Gardner:2015wea}).
Since QCD vector mesons are photoproduced copiously and exclusively, such reactions may provide an ideal laboratory in which to search for leptophobic bosons that couple to quarks.  

Searches for long-range nuclear forces and dark-photon decays to $e^+e^-$ place strong constraints on $B$ bosons with $m_B \lesssim m_{\pi}$\,\cite{Tulin:2014tya}. 
As we will show below, calculations for $m_B \gtrsim 0.9\gev$ are highly uncertain at the moment; therefore, we will focus here on the region $m_{\pi} \lesssim m_B \lesssim 0.9\gev$. 
Within this region, Refs.~\cite{Tulin:2014tya,JEF} showed that searches for $\eta \to B \gamma$ 
 are the most sensitive for $m_B \lesssim 0.5\gev$, while currently $\Upsilon \to$~hadrons provides the best limit at larger $m_B$.

In this article, we perform the first study of exclusive $B$ boson photoproduction, {\em i.e.} $\sigma(\gamma p \to p B)$. 
We will show that  the photoproduction data set to be collected by the {\sc GlueX} experiment at Jefferson Lab -- starting this year --  will provide the most sensitive probe of $B$ bosons with $m_B \gtrsim 0.5\gev$.  
The electroproduction data collected by the CLAS experiment at Jefferson Lab is also expected to be highly sensitive to $B$ bosons. 
Finally, while we focus on the $B$ boson, we stress that this search  will provide sensitivity to other proposed dark-sector states that couple to quarks.  

\section{Baryonic Force Model}

Following Ref.~\cite{Tulin:2014tya}, we consider the $B$ boson interaction Lagrangian
\begin{equation}
\label{eq:blag} 
\mathcal{L}_{\rm int} = \left(\frac{g_B}{3} + \varepsilon e_q\right) \bar{q}\gamma^{\mu} q B_{\mu} - \varepsilon e \bar{\ell} \gamma^{\mu} \ell B_{\mu},
\end{equation}
where $B_{\mu}$ is the new gauge field with universal gauge coupling to quarks $g_B$, $q$ are quark fields, $\ell$ are charged-lepton fields,  $e$ is the electromagnetic coupling, and $\varepsilon$ is the so-called kinetic mixing parameter between the $B$ and the photon. 
Assuming that a non-zero value of $\varepsilon$ arises at the one-loop level gives
\begin{equation}
\varepsilon = \mathcal{O}\left(\frac{e g_B}{(4\pi)^2}\right) \approx g_B \times 10^{-3}.
\end{equation}
The impact of the precise value of $\varepsilon$ on $B$ boson phenomenology in the mass region considered here is negligible, with the exception of the constraints derived from dark-photon searches.  

We make the following important observations about the baryonic dark force based on $\mathcal{L}_{\rm int}$:
\begin{itemize}
\item it preserves the low-energy symmetries of QCD;
\item it preserves $SU(3)$ flavor symmetry due to its universal quark coupling;
\item the $B$ boson has quantum numbers $I^G(J^{PC}) = 0^-(1^{--})$.
\end{itemize}
Since the $B$ boson has the same quantum numbers as the $\omega$ meson, in the $m_B$ region considered in this paper one expects the dominant $B$ decay modes to be $B \to \pi^+\pi^-\pi^0$ and $B \to \pi^0\gamma$.  
Indeed, as shown in Ref.~\cite{Tulin:2014tya}, $B \to \pi^0\gamma$ is dominant for $m_{\pi} \lesssim m_B \lesssim 0.6\gev$, while  $B \to \pi^+\pi^-\pi^0$ is dominant for $0.6\gev \lesssim m_B \lesssim 0.9\gev$.  
The only other decay mode with a branching fraction $\gtrsim 1\%$ in these $m_B$ regions is $B\to \pi^+\pi^-$ which is $\lesssim 5\%$ for all $m_B$ considered in this study. 
We note that for $m_B \gtrsim m_{\phi}$ the decay $B\to K^+K^-$ is expected to be important as well, 
since the $B$ boson also has the same quantum numbers as the $\phi$ meson.

\section{Photoproduction}

We calculate exclusive photoproduction of $B$ bosons within the hidden local symmetries (HLS) framework of vector meson dominance (VMD)~\cite{Bando:1984ej,Bando:1985rf,Bando:1987br,Fujiwara:1984mp}.
This framework is highly successful at predicting low-energy SM observables, which motivates its use here.
Within HLS-VMD, external gauge fields mix with the QCD vector mesons ($V = \rho,\omega,\phi,\ldots$).
For example,  the decay $\omega \to \pi^0\gamma$ is dominated by $\omega \to \pi^0\rho$ followed by $\rho \to \gamma$ mixing in the HLS-VMD framework.

To calculate the photoproduction cross section of $B$ bosons, we first consider the $V$ meson photoproduction amplitude
\begin{equation}
\mathcal{A}(\gamma p \to p V)^{\mu} V^*_{\mu},
\end{equation}
where for simplicity we suppress the photon and proton spin states, and leave the kinematic dependence implicit. 
Next, we introduce $V \to B$ mixing assuming a narrow $B$ boson,
which is done in a similar way as $V \to \gamma$ mixing but making the substitution~\cite{Tulin:2014tya}
\begin{equation}
e{\rm Tr}\left[T_V Q\right]  \to \frac{g_B}{3} {\rm Tr}\left[T_V\right],
\end{equation}
where $T_V$ are the $U(3)$ generators for $V$ (discussed below) and 
\begin{equation}
Q = \frac{1}{3}
\begin{bmatrix}
  2 & 0 & 0\\
  0 & -1 & 0\\
  0 & 0 & -1 
\end{bmatrix}
\end{equation}
gives the coupling of the $\gamma$ to quarks. Since the $B$ has universal quark couplings, $Q$ is replaced by the identity matrix for the $B$ boson. 
The amplitude for $\gamma p \to p (V\to B)$ is then 
\begin{equation}
\label{eq:amp}
\mathcal{A}(\gamma p \to\! p V)^{\mu} B^*_{\mu} \left[\frac{2 g_B m_V^2 {\rm Tr}{\left[T_V\right]}}{3 \sqrt{12\pi} D_V(m_B)} \right],
\end{equation}
where $B_{\mu}$ is the physical $B$ boson polarization vector with $p_B \cdot B = 0$, 
the VMD gauge coupling is extracted from $\Gamma(\rho\to\pi\pi)$ to be $\approx\sqrt{12\pi}$~\cite{Fujiwara:1984mp}, 
\begin{equation}
D_V(m_B) = m_V^2 - m_B^2 - i m_B \Gamma_V(m_B),
\end{equation}
and $\Gamma_V(m)$ is the mass-dependent width of $V$ taken from Ref.~\cite{Achasov:2003ir}.  
We note that $\mathcal{A}(\gamma p \to\! p V)$ should be evaluated for $p B$ kinematics, which are different from $p V$ since $m_B \neq m_V$; however, $|\vec{p}_B|_{\rm c.m.}$, $t$, and $u$ have little dependence on $m_B$ in the mass range considered here for the {\sc GlueX} photon-beam energy (and $s$ is approximately constant).
Therefore, the $m_B$ dependence of the amplitude is expected to be small.
Once the data are collected, the mass dependence of the strength of the $0^-(1^{--})$ partial wave can be used to confirm this, or to derive a correction factor if one is required.

We do not consider $B\to\rho$ mixing as this violates isospin conservation.  The relevant $U(3)$ generators for the $\omega$ and $\phi$ mesons are 
\begin{equation}
T_{\omega} = \frac{1}{2}
\begin{bmatrix}
  1 & 0 & 0\\
  0 & 1 & 0\\
  0 & 0 & 0 
\end{bmatrix}, \qquad
T_{\phi} = \frac{1}{\sqrt{2}} 
\begin{bmatrix}
  0 & 0 & 0\\
  0 & 0 & 0\\
  0 & 0 & 1 
\end{bmatrix},
\end{equation}
\\
\noindent assuming $\omega = (u\bar{u}+d\bar{d})/\sqrt{2}$ and  $\phi = s\bar{s}$. 
We neglect both isospin and $SU(3)$ breaking effects since both are known to be small for the $\rho$, $\omega$, and $\phi$ mesons. 
The $B$ boson photoproduction amplitude is then
\begin{eqnarray}
\mathcal{A}(\gamma p \to p B) \approx \left[\frac{2 g_B}{3 \sqrt{12\pi}}\right] \hspace{1.4in} \nonumber \\ 
\times \left( \frac{m_{\omega}^2\mathcal{A}(\gamma p \to p \omega)^{\mu}}{D_{\omega}(m_B)} +  \frac{m_{\phi}^2 \mathcal{A}(\gamma p \to p \phi)^{\mu}}{\sqrt{2} D_{\phi}(m_B)} \right) B^*_{\mu}.
\end{eqnarray}
Therefore, if $\mathcal{A}(\gamma p \to p V)$ can be determined from the data for $V = (\omega,\phi)$, the $B$ boson production rate can be inferred for a given $g_B$.  

\begin{figure}[]
\includegraphics[width=0.8\columnwidth]{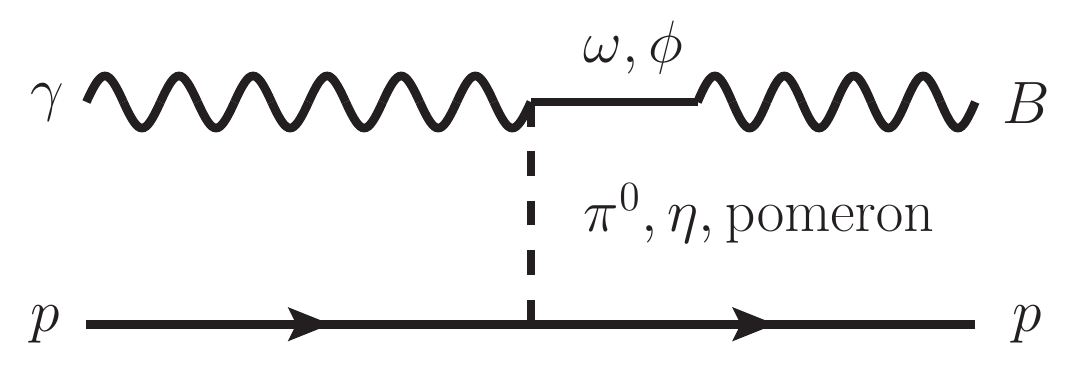}
\caption{
\label{fig:feynman}
Feynman diagrams for photoproduction of $B$ bosons. At {\sc GlueX} energies, pion and pomeron exchange are expected to be dominant for $\omega$ photoproduction, while $\phi$ photoproduction is expected to be dominantly diffractive.  
These diagrams are shown for illustrative purposes only; no model of $\omega$ or $\phi$ photoproduction is assumed in this work.  
}
\end{figure}

Our study targets the {\sc GlueX} experiment at Jefferson Lab which will employ a linearly polarized photon beam with $E_{\gamma} \approx 8-9\gev$ corresponding to 
\begin{equation}
s = m_p^2 + 2m_pE_{\gamma} \approx (4\gev)^2.
\end{equation}
The most relevant Feynman diagrams for $V$ photoproduction at this energy are shown in Fig.~\ref{fig:feynman}. 
Photoproduction processes that induce a spin flip of the target proton add incoherently with those that do not.
For example, $t$-channel pion exchange amplitudes do not interfere with those for diffractive production.  
These processes are often referred to as {\em unnatural} and {\em natural} parity exchange, respectively, and we will adopt the notation of $-$ and $+$ to correspond with the parity of the pion and pomeron.
We stress, however, that our results are not model dependent, since no assumption is made about the structure of the spin (non)flip amplitudes.

The total $B$ boson photoproduction cross section is obtained from the sum of the natural and unnatural components
\begin{equation}
\sigma(\gamma p \to p B) = \sigma_+(\gamma p \to p B) + \sigma_-(\gamma p \to p B),
\end{equation}
where the $B$ boson cross sections are related to those of the $\omega$ and $\phi$ by
\begin{widetext}
\[
\label{eq:main}
\sigma_{\pm}(\gamma p \!\to\! p B) \approx \frac{4 \alpha_B}{27} \Phi(m_B)\!\! \left[ \frac{m^4_{\omega} \sigma_{\pm}(\gamma p \!\to\! p \omega)}{\Phi(m_{\omega}) |D_{\omega}(m_B)|^2} \!+\! \frac{m^4_{\phi} \sigma_{\pm}(\gamma p \!\to\! p \phi)}{2 \Phi(m_{\phi}) |D_{\phi}(m_B)|^2} \!+\!  \frac{\sqrt{2}\cos{(\varphi_{\pm})}m_{\omega}^2m^2_{\phi}\sqrt{ \sigma_{\pm}(\gamma p \!\to\! p \omega)  \sigma_{\pm}(\gamma p \!\to\! p \phi)}}{|D_{\omega}(m_B)| |D_{\phi}(m_B)|  \sqrt{\Phi(m_{\omega}) \Phi(m_{\phi})}}\right]\!.
\]
\end{widetext}
Here $\alpha_B \equiv g_B^2/4\pi$, $\varphi_{\pm}$ is the phase difference between the $\omega$ and $\phi$ amplitudes (with implicit $s$, $t$, $m_B$ dependence), 
and 
\begin{equation}
\frac{\Phi(m_B)}{\Phi(m_V)} = \frac{\sqrt{(s - (m_p+m_B)^2) (s - (m_p - m_B)^2)}}{\sqrt{(s - (m_p+m_V)^2) (s - (m_p - m_V)^2)}}
\end{equation}
are the usual phase-space factors. In the $m_B$ range studied here and at {\sc GlueX} energies 
\begin{equation}
\frac{\Phi(m_B)}{\Phi(m_V)} \approx 1. 
\end{equation}
The values of $\sigma(\gamma p \to p V)$ can be measured in the same data set in which the $B$ boson is to be searched for.  Furthermore, since {\sc GlueX} will employ a linearly polarized photon beam, $\sigma_{\pm}(\gamma p \to p V)$ can be obtained by measuring both $\sigma(\gamma p \to p V)$ and the $V$ spin density matrix elements.  
We expect that such measurements will be made even in the absence of a $B$ boson search, so the only missing input to infer the $B$ boson signal in a data-driven manner is $\varphi_{\pm}$.

\begin{figure}[]
\includegraphics[width=0.99\columnwidth]{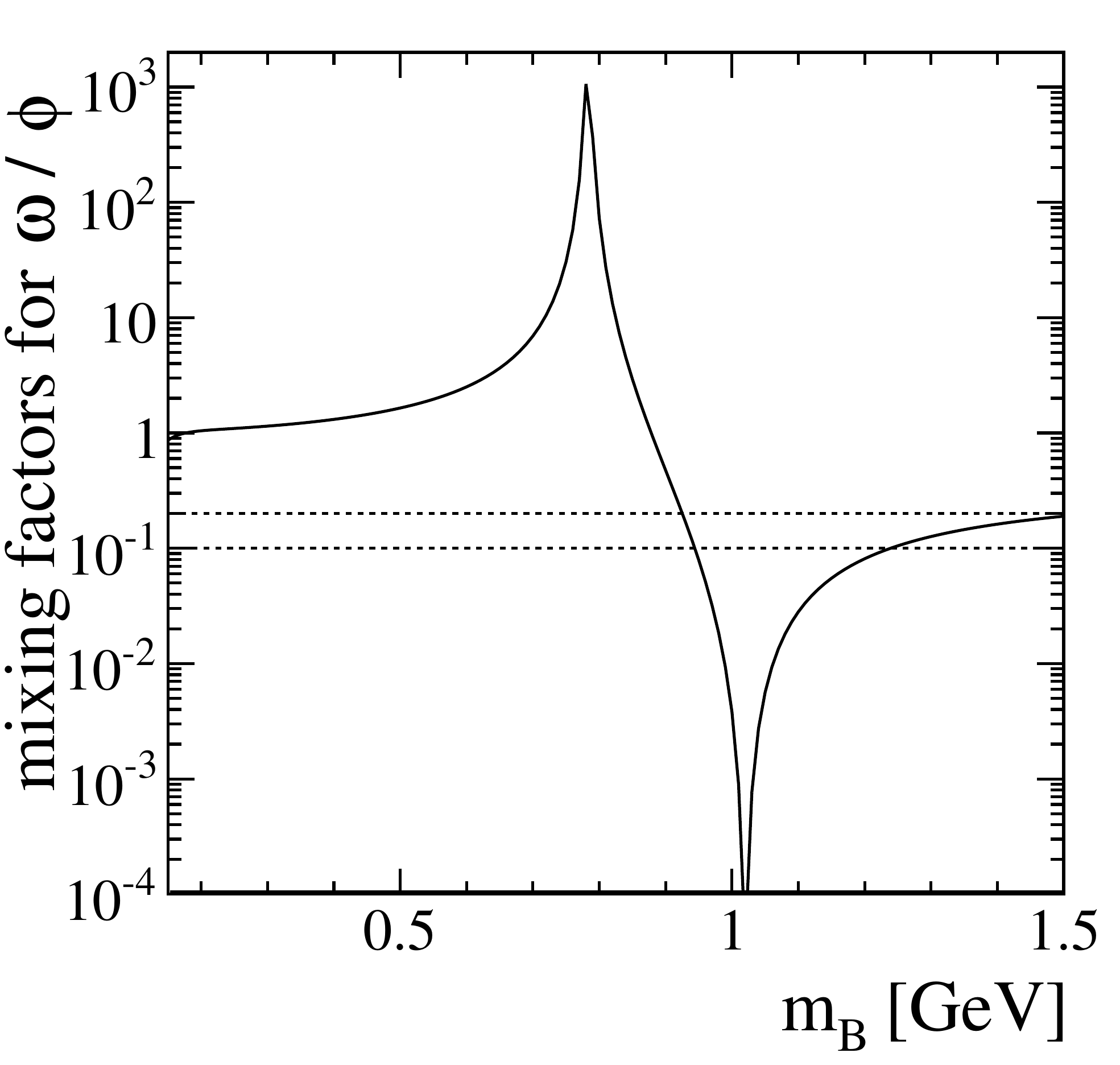}
\caption{
\label{fig:fofp}
When $\mathcal{R}^{\omega}_{\phi}  \equiv | \frac{m_{\omega}^2}{D_{\omega}(m_B)}|^2 / | \frac{m_{\phi}^2}{D_{\phi}(m_B)}|^2$  has a value that falls approximately between the horizontal dashed lines the natural $\omega$ and $\phi$ contributions to $B$ boson photoproduction will cancel due to destructive interference if $\varphi_+ \approx \pi$. 
}
\end{figure}

Existing $\omega$ and $\phi$ photoproduction measurements suggest the following for $s \approx (4\gev)^2$~\cite{PhysRevD.7.3150}:
\begin{eqnarray}
\frac{\sigma_+(\gamma p \to p \omega)}{\sigma_+(\gamma p \to p \phi)} &\approx& 4, \nonumber \\ 
\frac{\sigma_-(\gamma p \to p \omega)}{\sigma_+(\gamma p \to p \omega)} &\approx& 0.1, \\ 
\frac{\sigma_-(\gamma p \to p \phi)}{\sigma_+(\gamma p \to p \phi)} &\approx& 0. \nonumber
\end{eqnarray}
Assuming the high-precision measurements that will be made at {\sc GlueX} confirm this hierarchy, the unnatural component is simply
\begin{equation}
\frac{\sigma_{-}(\gamma p \!\to\! p B)}{\sigma_{-}(\gamma p \!\to\! p \omega)} \approx \frac{4 \alpha_B m^4_{\omega} }{27 |D_{\omega}(m_B)|^2}, 
\end{equation}
while the natural term is 
\begin{eqnarray}
\frac{\sigma_+(\gamma p \to p B)}{\sigma_+(\gamma p \to p \omega)} &\approx& \frac{4 \alpha_B}{27} \Big[ \left| \frac{m_{\omega}^2}{D_{\omega}(m_B)}\right|^2 + \frac{1}{8} \left| \frac{m_{\phi}^2}{D_{\phi}(m_B)}\right|^2 \nonumber \\ &+& \frac{\cos{(\varphi_+)}m_{\omega}^2m^2_{\phi}}{\sqrt{2} |D_{\omega}(m_B)| |D_{\phi}(m_B)|} \Big]. 
\end{eqnarray}
Figure~\ref{fig:fofp} shows the $m_B$ dependence of the ratio of $\omega$ to $\phi$ mixing factors
\begin{equation}
\mathcal{R}^{\omega}_{\phi} \equiv \frac{\left| \frac{m_{\omega}^2}{D_{\omega}(m_B)}\right|^2}{ \left| \frac{m_{\phi}^2}{D_{\phi}(m_B)}\right|^2} \,\,.
\end{equation}
As expected, for $m_B \approx m_{\omega}$ we can ignore mixing with the $\phi$, whereas
for $m_B \lesssim 0.5\gev$ we find
\begin{equation}
\frac{\sigma_+(\gamma p \!\to\! p B)}{\sigma_+(\gamma p \!\to\! p \omega)} \approx \frac{\alpha_B}{6 \left(1 - \frac{m^2_B}{m^2_{\omega}}\right)^2} \left[1 + 0.5\cos{(\varphi_+)}\right].
\end{equation}
Therefore, without any constraints on $\varphi_+$ the uncertainty in the inferred $B$ boson natural cross section is less than a factor of two for $m_B \lesssim 0.5\gev$.  

For $m_B \gtrsim 0.9\gev$ -- and away from $m_{\phi}$ -- the natural $\omega$ and $\phi$ contributions to $B$ boson photoproduction could cancel due to destructive interference, {\em i.e.}\  $\sigma_+(\gamma p \to p B) \approx 0$ if $\varphi_+ \approx \pi$.  
High-precision measurements of $\omega$ and $\phi$ photoproduction at {\sc GlueX} will determine:
\begin{itemize}
\item If $\sigma_-(\gamma p \to p \omega) \gg \sigma_-(\gamma p \to p \phi)$ as has been observed at lower $E_{\gamma}$, then $\varphi_-$ does not need to be known and a conservative sensitivity can be determined by assuming maximal destructive interference at each $m_B$; {\em i.e.}, one can assume that $\sigma_+(\gamma p \to p B) = 0$ and that only $\sigma_-(\gamma p \to p B)$ contributes.
\item If $\varphi_+$ can be determined in a data-driven manner, then inferring the sensitivity within HLS-VMD is straightforward. 
In Sec.~\ref{sec:improve}, we discuss the possibility of determining $\varphi_+$.
\item Alternatively, if the kinematic dependence of the spin density matrix elements can be described well by a diffractive model of $\gamma p \to p V$ for both the $\omega$ and $\phi$, then $\varphi_+$ could be calculated within the model. 
\end{itemize}  
Therefore, the situation will become much more clear once the data is collected.  For this study, we choose to truncate our sensitivity predictions at $m_B = 0.9\gev$, but stress that it is possible the sensitivity at higher masses can be inferred using HLS-VMD.

\section{Sensitivity}

We estimate the sensitivity for the {\sc GlueX} experiment for its so-called Phase~IV run period, which is expected to begin in 2017 and last for about two years~\cite{GLUEX}. We take the total number of photons with $8 \lesssim E_{\gamma} \lesssim 9\gev$ on target to be
\begin{equation}
N_{\gamma} \approx 7\times 10^{14},
\end{equation} 
which takes the total Phase~IV run time and assumes an 80\% live time.  
{\sc GlueX} employs a 30-cm-thick liquid hydrogen target so
\begin{equation}
N_p/{\rm cm}^2 = 1.28 \times 10^{24}.
\end{equation}
Finally, using previous measurements of $\omega$ photoproduction we estimate~\cite{PhysRevD.7.3150}
\begin{equation}
\sigma(\gamma p \to p \omega) \approx 2\mu{\rm b},
\end{equation}
which gives
\begin{equation}
N(\gamma p \to p \omega) \approx 2 \times 10^9,
\end{equation}
excluding reconstruction and selection efficiencies.

\begin{figure}[]
\includegraphics[width=0.99\columnwidth]{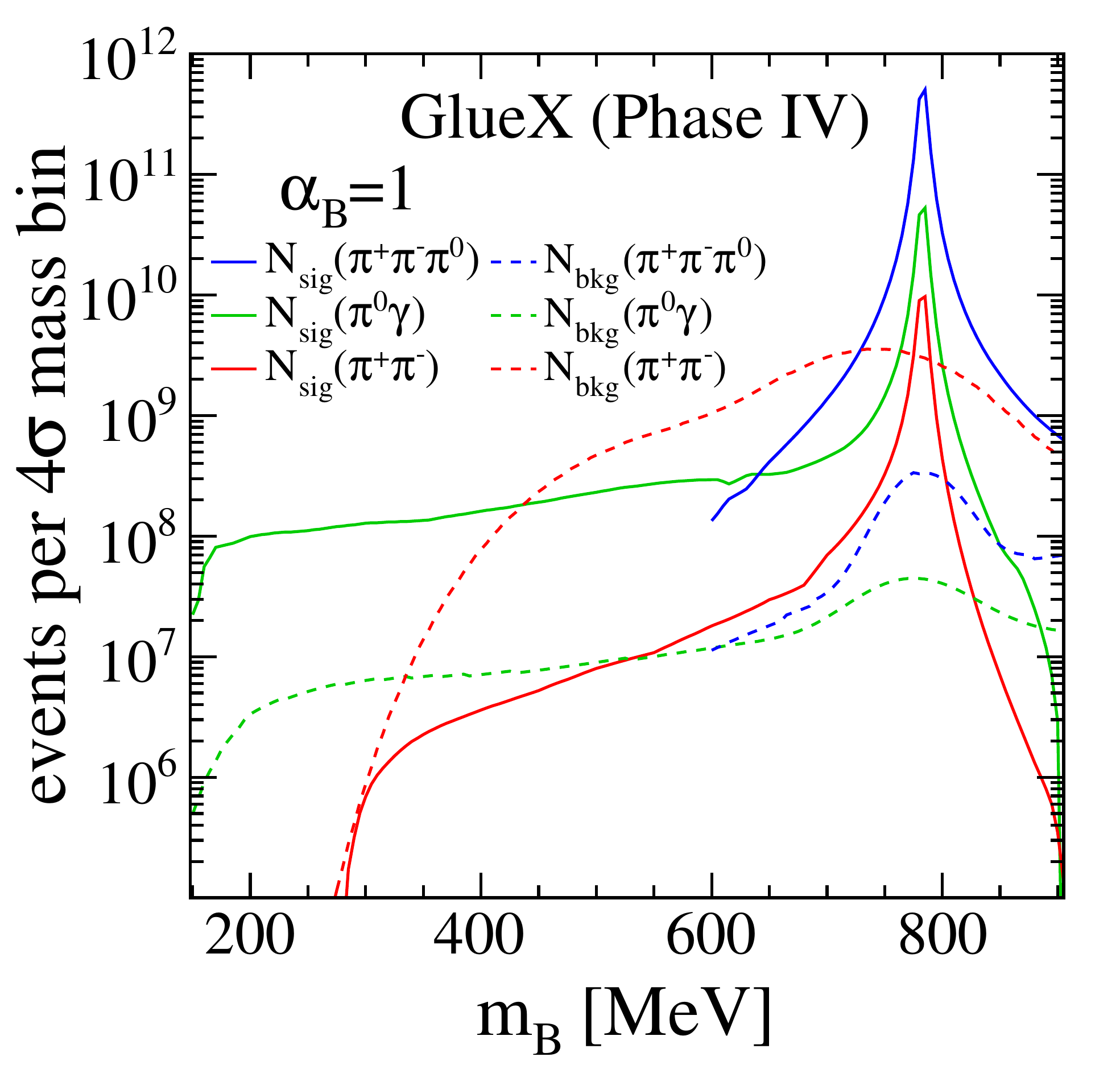}
\caption{
\label{fig:yields}
Expected reconstructed and selected $B$ boson yields versus $m_B$ for $\alpha_B = 1$.  Signal yields scale linearly with $\alpha_B$. 
The expected background yields are also shown. 
{\em N.b.}, we do not consider the $\pi^+\pi^-\pi^0$ final state below 0.6\gev to avoid the $\eta$ mass peak. 
}
\end{figure}

Our nominal search involves the decays $B \to \pi^0\gamma$, $B \to \pi^+\pi^-\pi^0$, and $B \to \pi^+\pi^-$.
We estimate the efficiency $\epsilon$ to reconstruct and select the reaction $\gamma p \to p B$ for these $B$ decay modes using the official {\sc GlueX} {\sc Geant}-based simulation package. 
For concreteness, we assume the $\omega$ photoproduction model of Ref.~\cite{Oh:2000zi} as it describes existing data at lower energies well~\cite{Williams:2009ab}; however, due to the hermetic nature of the {\sc GlueX} detector the dependence of the acceptance on the $\omega$ photoproduction model chosen is small.  
We stress that in the actual $B$ boson search, no model-dependent assumptions need to be made provided that both the search and normalization measurements are performed in the same fiducial region. 
For each mode, the efficiency is found to be $\epsilon \approx 20-30\%$ except near threshold. 
The expected $B$ boson yield in the final state $X$ is taken to be
\begin{eqnarray}
\label{eq:byield} 
N(\gamma p \to p B \to p X) \approx  \hspace{1.5in} \nonumber \\ \frac{4 \alpha_B m^4_{\omega} }{27 |D_{\omega}(m_B)|^2} N(\gamma p \to p \omega) \epsilon (p X) \mathcal{B}(B \to X). 
\end{eqnarray}
Figure~\ref{fig:yields} shows the expected reconstructed and selected $B$ boson yield for $\alpha_B = 1$ in each decay mode considered in our study.

The official {\sc GlueX} simulation package is also used to determine the expected $m_B$ resolution:
\begin{eqnarray}
\frac{\sigma(m_{\pi\gamma})}{m_{\pi\gamma}} &\approx&  2-3\%,\nonumber \\
\frac{\sigma(m_{3\pi})}{m_{3\pi}} &\approx&  1-2\%, \\
\frac{\sigma(m_{2\pi})}{m_{2\pi}} &\approx&  3-4\%,
\end{eqnarray}
where we perform a kinematic fit that utilizes the well-known initial state 4-momentum to improve the resolution on the final-state kinematics.

Estimating the background yields is difficult since many photoproduction processes can result in either a $p\pi^0\gamma$, $p\pi^+\pi^-\pi^0$, or $p\pi^+\pi^-$ final state. 
For the $3\pi$ and $2\pi$ modes, we take the background spectra from a previous $\gamma p \to p V$ study at $E_{\gamma} = 9.3\gev$~\cite{PhysRevD.7.3150}, and assume that any reducible mis-reconstructed contributions in those spectra are the same at {\sc GlueX}.  
For the $\gamma p \to p \pi^+\pi^-$ reaction, the background is dominantly from $\gamma p \to p \rho$.  
We are not aware of any existing measurements of $\gamma p \to p \pi^0\gamma$ using high-energy photons. 
To estimate the $m(\pi\gamma)$ spectrum, we first note that the $m(3\pi)$ spectrum published by {\sc GlueX} using lower-energy $E_{\gamma} \approx 3\gev$ photons~\cite{GLUEXMASS} is similar in shape and in size relative to the $\omega$ yield as that of Ref.~\cite{PhysRevD.7.3150}. 
Therefore, we take the shape and size relative to the $\omega$ peak of the $m(\pi\gamma)$ spectrum from Ref.~\cite{GLUEXMASS}, and simply rescale this to the expected $\omega$ yield for $E_{\gamma} = 8-9\gev$.  

While we have checked that {\sc Pythia}~\cite{Sjostrand:2014zea} also suggests the background shapes at $E_{\gamma} = 3$ and 9.3\gev are similar, and that the background yields relative to that of the $\omega$ are approximately constant with $E_{\gamma}$, one should view our backgrounds as approximate.
In the actual $B$ boson search, the background yields for each $m_B$ can be estimated using the data sidebands.  
Therefore, while the uncertainty in predicting the background yields here results in an uncertainty on the potential sensitivity to $B$ of a factor of $\approx 3$, the impact on the actual sensitivity will be governed by the precision with which one can interpolate the background yield using the data sidebands which is expected to be a small effect.  

\begin{figure}[]
\includegraphics[width=0.99\columnwidth]{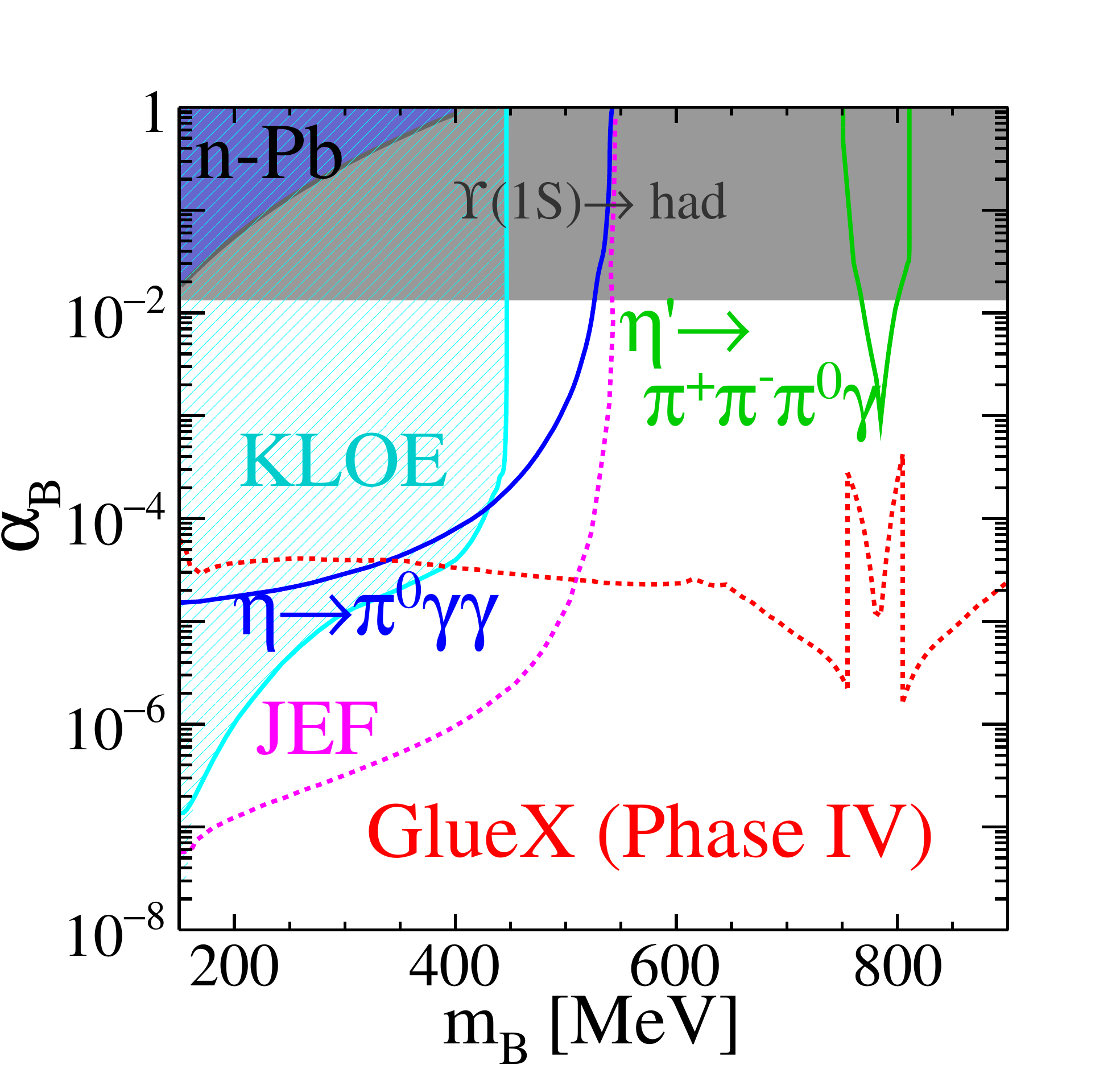}
\caption{
\label{fig:limits}
Existing limits on $B$ bosons~\cite{Tulin:2014tya} compared to the expected sensitivity obtained from $\eta$ meson decays at JEF~\cite{JEF} and photoproduction (this work).
Only the $B \to \pi^+\pi^-$ decay is used near $m_{\omega}$ in our study, since the large $\omega$ peak in $\pi^0\gamma$ and $\pi^+\pi^-\pi^0$ will complicate searching for a $B$ boson with similar mass.  
The existing limits are derived from low-energy n-Pb scattering~\cite{Barbieri:1975xy}, the total rate of $\Upsilon$ decays into hadrons~\cite{Aranda:1998fr}, and decays of $\eta$ and $\eta'$ mesons. 
The KLOE~\cite{Babusci:2012cr} sensitivity depends on the kinetic mixing parameter with the photon and is displayed here assuming $\varepsilon = e g_B /(4\pi)^2$. 
}
\end{figure}

To estimate the sensitivity at {\sc GlueX} to $B$ boson photoproduction, we use as a rough criterion for the exclusion limits
\begin{equation}
\frac{N(\gamma p \to p B \to p X; m_B, \alpha_B)}{\sqrt{N(\gamma p \to p X; m_B)}} \approx 2,
\end{equation}
where the yields are in the region $m_B - 2\sigma(m) < m < m_B + 2\sigma(m)$.
Figure~\ref{fig:limits} shows that using $B$ boson photoproduction one can probe $\alpha_B$ values down to  $\mathcal{O}(10^{-5})$, making this the most sensitive probe for $m_B \gtrsim 0.5\gev$.   
Finally, we note that while we considered the full Phase~IV data set, it is clear that {\sc GlueX} will be able to probe unexplored $B$ boson parameter space with a much smaller data set, {\em e.g.}, the data to be collected later this year. 

The sensitivity of dark-photon searches to $B$ bosons is model dependent in part because $\mathcal{B}(A' \to \ell^+\ell^-)$, where $\ell \equiv e$ or $\mu$, depends on the kinetic-mixing parameter $\varepsilon$.  
Figure~\ref{fig:limits} shows the exclusion region obtained from a KLOE~\cite{Babusci:2012cr} dark-photon search for the decay $\phi \to \eta A'(e^+e^-)$, where here we assume $\varepsilon = e g_B /(4\pi)^2$. 
This decay has good sensitivity to $B$ bosons due to the large predicted value of $\mathcal{B}(\phi \to \eta B)$\cite{Tulin:2014tya}.
Searches for $e^+e^- \to \gamma A'(\ell^+\ell^-)$ do not provide relevant constraints on $\alpha_B$ since both the production and decay of the $B$ boson rely on its small leptonic coupling.  

\section{Potential Improvements}
\label{sec:improve}

There are a number of potential improvements to this search.  
The non-$V$ background may have a different angular distribution than a vector boson produced by a polarized photon beam and this could be exploited to improve the sensitivity. 
The proposed upgrade of the {\sc GlueX} forward calorimeter as part of the so-called JEF program would improve the invariant-mass resolution~\cite{JEF}. 
{\sc GlueX} may collect more integrated luminosity than we assumed here as it has been approved for additional running that will include an upgraded particle-identification system~\cite{Dugger:2014xaa}. 

In principle, the $\omega-\phi$ phase differences $\varphi_{\pm}$ could be determined by studying the $J^P = 1^{--}$ partial wave strength versus $m(3\pi)$ in the $p\pi^+\pi^-\pi^0$ final state, or possibly versus $m(ee)$ in $\gamma p \to p e^+ e^-$. 
This would greatly reduce the uncertainty in the normalization due to interference between the $\omega$ and $\phi$ amplitudes.  
Indeed, a dip consistent with $\omega-\phi$ interference is seen in $e^+e^- \to \pi^+\pi^-\pi^0$ which determines the phase difference to be near the expected value based on $SU(3)$ including a small $\omega-\phi$ mixing correction~\cite{Achasov:2003ir}.  
That said, if the natural and unnatural $\omega$ and $\phi$ cross sections are as predicted in Ref.~\cite{Oh:2000zi,Titov:2003bk}, then the normalization uncertainty for $m_B \lesssim 0.9\gev$ due to this unknown phase is at most $\approx 30\%$. 
At this level, one should view the use of the HLS-VMD framework itself with some degree of skepticism.  
For the purpose of setting exclusions, the use of HLS-VMD and ignoring the $\phi$ interference  should be sufficient. 
As discussed previously, at higher masses $\varphi_+$ must be measured to determine the sensitivity using HLS-VMD, and
the decay $B \to K^+K^-$ should be included in the search.  
Finally, in the case of a discovery, we note that lattice QCD should be able to provide more precise results than HLS-VMD. 

One could consider searching for displaced $B$ boson decays which would have considerably less background contamination. 
The resolution on the $B$ flight distance, {\em i.e.}\ the experimental resolution on the distance between the production and decay positions of the $B$ boson, 
in the $2\pi$ and $3\pi$ decay modes is expected to be $\approx 1$\,cm. 
We do not expect {\sc GlueX} to collect sufficient luminosity to do a displaced search for $B$ bosons in hadronic decay modes; 
however, we encourage pursuing a displaced search as it may be sensitive to unexplored parameter space of more general dark-sector models.   

\section{Electroproduction}

Our results also motivate looking for electroproduction  of $B$ bosons at the CLAS experiment at Jefferson Lab. 
Vector meson electroproduction dominantly proceeds via an off-shell photon $\gamma^*$.  
The $\gamma^*$ kinematics vary event-by-event, making it possible for CLAS to include only regions of phase space where
\begin{equation}
\sigma(e p \to e p \omega) \gg \sigma(e p \to e p \phi),
\end{equation}
which includes most of the potential signal~\cite{Morand:2005ex,Santoro:2008ai}.  
This enables considering only $B\to\omega$ mixing, where the expected $B$ boson electroproduction yield is given by Eq.~\ref{eq:byield} with photoproduction reactions replaced by the corresponding electroproduction ones.

\section{Other Dark-Sector Theories}

More general models are also possible; {\em e.g.}, Eq.~\ref{eq:blag} could be modified to have non-universal quark couplings.   
In this case, the new boson $B'$ need not be an isoscalar and mixing with the $\rho$ would also be possible.  
Without fully specifying the quark couplings $g_{B'}^{q}$ of the model, we cannot repeat the exercise of mapping $B'$ yields directly to $g_{B'}^{q}$.  
Therefore, we suggest that {\sc GlueX} measure -- or set upper limits on -- each of the ratios
\begin{equation}
\frac{\sigma(\gamma p \to p B')\times \mathcal{B}(B' \to X)}{\sigma(\gamma p \to p V)}, 
\end{equation}
where $X = (\pi^0\gamma, 2\pi, 3\pi, K^+K^-, K^*K, \ldots)$ and $V = (\rho, \omega, \phi)$, while scanning in $m_{B'}$ and $B'$ lifetime. 
There is no reason to truncate the search at any $m_B$. 
The results of these measurements can subsequently be recast to determine the sensitivity to models that predict new bosons that couple to quarks.  
See Ref.~\cite{Gardner:2015wea} for an expanded discussion on such theories. 

\section{Summary}

In summary, we proposed a search for exclusive photoproduction of a gauge boson that couples to baryon number $B$ at the {\sc GlueX} experiment at Jefferson Lab. 
We determined that $\gamma p \to p B$ will provide the best sensitivity for $B$ boson masses above 0.5\gev. 
This search will also provide sensitivity to other proposed dark-sector states that couple to quarks.  
Finally, our results motivate a similar search for $B$ boson electroproduction at the CLAS experiment.

\section*{Acknowledgements}

We thank the {\sc GlueX} collaboration for permitting us to use their official simulation package; and  
 W.~Detmold, L.~Gan, P.~Ilten, D.~Mack, Y.~Soreq, J.~Stevens, J.~Thaler, S.~Tulin, and W.~Xue
for useful discussions and feedback.
This work was supported
by US Department of Energy (DOE) grant DE-SC0010497.

\bibliography{bib}

\end{document}